\documentclass[a4paper]{spie}  
\usepackage[]{graphicx}
\usepackage[left=1.925cm, right=1.925cm, top=2.54cm, bottom=4.94cm]{geometry}
\title{Background estimation in a wide-field background-limited instrument such as Fermi GBM}

\author{Gerard Fitzpatrick\supit{a}, 
        Sheila McBreen\supit{a}, 
		Valerie Connaughton\supit{b} and
		Michael Briggs\supit{b} 
		on behalf of the GBM collaboration\footnote{\hspace{1ex}http://gammaray.nsstc.nasa.gov/gbm/people/collaborators/}
\skiplinehalf
\supit{a}School of Physics, University College Dublin, Belfield, Dublin 4, Ireland; \\
\supit{b}University of Alabama in Huntsville, NSSTC, 320 Sparkman Drive, Huntsville, AL 35805, USA
}

\authorinfo{Further author information:\\G. Fitzpatrick: E-mail: gerard.fitzpatrick@ucdconnect.ie, Telephone: +353 1716 2202}

\begin{document} 
\maketitle 

\begin{abstract}
The supporting instrument on board the \textit{Fermi} Gamma-ray Space Telescope, the Gamma-ray Burst Monitor (GBM) is a wide-field gamma-ray monitor composed of 14 individual scintillation detectors, with a field of view which encompasses the entire unocculted sky. Primarily designed as transient monitors, the conventional method for background determination with GBM-like instruments is to time interpolate intervals before and after the source as a polynomial. This is generally sufficient for sharp impulsive phenomena such as Gamma-Ray Bursts (GRBs) which are characterised by impulsive peaks with sharp rises, often highly structured, and  easily distinguishable against instrumental backgrounds. However, smoother long lived emission, such as observed in solar flares and some GRBs, would be difficult to detect in a background-limited instrument using this method. We present here a description of a technique which uses the rates from adjacent days when the satellite has approximately the same geographical footprint to  distinguish low-level emission from the instrumental background. We present results from the application of this technique to GBM data and discuss the implementation of it in a generalised background limited detector in a non-equatorial orbit.
\end{abstract}

\section{INTRODUCTION}
\label{sec:intro} 
Wide-field instruments have played an important role in the development of gamma-ray astronomy by performing the important task of simultaneously monitoring a large fraction of the sky. They can therefore alert highly sensitive, narrow-field instruments to transient activity that would otherwise not be detected. An important example is the detection of the first Gamma-Ray Burst (GRB) afterglow in the optical and X-ray regimes~\cite{paradijs_97, costa_97}. Scintillation detectors in the gamma-ray regime are particularly well suited to the role of monitors, due to their wide field of view, large energy bandwidth and affordability (relative to more complex focused detectors).

Such systems have several limitations, primarily the lack of positional information. This can be somewhat ameliorated by using several detectors and comparing the relative rates in each to derive the approximate location. This typically gives locations with accuracy on the order of degrees. A further limitation is that in general, such systems are optimised for the study of bright impulsive events, cannot able to easily separate smooth emission from the background.

For satellites in low earth orbits, the instrumental background has large variations over the course of an orbit due to the variation of local particle flux densities in the atmosphere. A secondary effect is activation of the spacecraft materials following passage through the South Atlantic Anomaly (SAA). The relatively short periods of such orbits also means that bright sources (e.g. Crab, Cyg X-1 and Sco X-1) which enter/exit earth occultation will result in a step in the background continuum. In order to make accurate measurements, a confident estimation of this variable background level is necessary. The conventional method for determining the background in wide-field instruments is to interpolate time intervals before and after the event of interest as a polynomial (usually of order 0-4). This is particularly well suited to the study of transient phenomena such as GRBs, the prompt emission of which is characterized by impulsive peaks with sharp rises, often highly structured, and easily distinguishable against instrumental backgrounds. The timescales on which the prompt emission occurs is usually short enough that this method is sufficient, however for events with long durations or less impulsive time profiles, a more rigorous background determination is required. 

In this work we present one method of background estimation which uses the rates from alternate days when the satellite had the same geographical footprint. In \S~\ref{sec:motivation} we present the scientific motivation/requirement for a more rigorous background determination and in \S~\ref{sec:fermi} we describe the \emph{Fermi} telescope and the Gamma-ray Burst Monitor (GBM). In \S~\ref{sec:os} we present a detailed description of our method of background estimation, and in \S~\ref{sec:verif} we discuss several ways in which our technique was validated.

\section{Scientific Motivation}
\label{sec:motivation}
Capable of making observations across an unprecedented 8 decades of energy (10 keV - 300 GeV), the \emph{Fermi} Gamma-Ray Space Telescope has ushered in a new era in gamma-ray astrophysics. This broad energy bandwidth is afforded by its two constituent instruments, the Large Area Telescope (LAT), a pair-production telescope with an effective energy bandwidth of 20 MeV - 300 GeV~\cite{lat_paper} and the supporting instrument, the Gamma-ray Burst Monitor (GBM), which consists of 14 individual scintillation detectors with an effective energy range of 10 keV - 40 MeV~\cite{gbm_paper}. Gamma-ray bursts (GRBs) and solar flares are two of the strong science drivers for both instruments on-board \emph{Fermi}.

\begin{figure} \begin{center} \begin{tabular}{c}
\includegraphics[scale = 0.3]{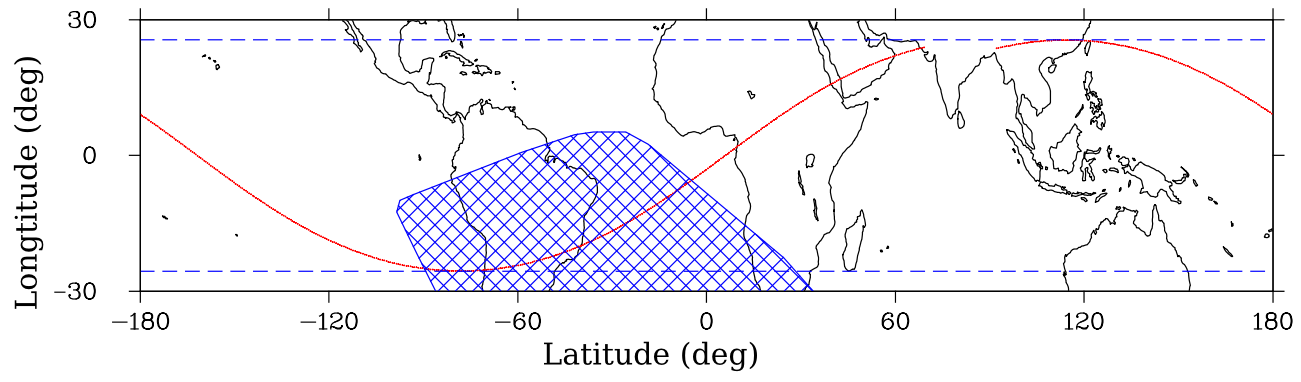}
\end{tabular} \end{center} \caption[example] 
{ \label{fig:orbit} The orbit of \emph{Fermi}: The hatched region is the South Atlantic Anomaly (SAA), the horizontal dashed lines indicate the limits of the inclination of \emph{Fermi} and the solid line is the path of the spacecraft over one orbit. } \end{figure} 

With GRBs, this has lead directly to the discovery of additional high and low energy spectral components in addition to the empirical `Band'  model~\cite{band_93} (e.g. Abdo et al.~\cite{2009ApJ...706L.138A} and Ackermann et al.~\cite{ackermann_11}). The sensitivity of the Large Area Telescope (LAT) on board \emph{Fermi} has also led to the discovery of long lived emission in the MeV-GeV energy range on kilo-second timescales. Extended emission from GRBs above 100 MeV had previously been hinted at by observations made by the EGRET instrument on board the Compton Gamma Ray Observatory~\cite{1994Natur.372..652H}. This emission has been observed in several GRBs by the LAT and can be explained as synchrotron emission from the external forward shock~\cite{ghisellini_10,kumar_10}. These observations on timescales which have been conventionally associated with lower energy (sub-MeV) afterglow emission have raised the question of whether such emission would be visible in GBM.

\emph{Fermi} GBM has detected 511 keV positron annihilation, 2.2 MeV neutron capture lines, and several other nuclear lines up to 7 MeV from an M-class solar flare on June 12 2010~\cite{ackermann_12_sfl}. The Large Area Telescope (LAT) detected emission above 100 MeV from this flare, most likely from pion decay. Current predictions place the sunspot number maximum during February 2013\footnote{http://solarscience.msfc.nasa.gov/predict.shtml}, and as it approaches we can expect the frequency of solar flares to increase. Class X and M flares can have durations on the order of hours, and in order to study them with GBM, and in particular to trace the evolution in time of the nuclear line component, a reliable background subtraction method is required. Once this is achieved, the evolution of the nuclear emission with the non-thermal emission from accelerated particles can be studied, at both GBM and LAT energies.

GBM is the successor of the highly successful BATSE instrument on board the Compton Gamma Ray Observatory. Connaughton (2002)\cite{connaughton_02} employed a method of background estimation to create background subtracted signals from hundreds of BATSE GRBs. By summing these, extended emission up to hundreds of seconds after the bursts was found. This provides clear evidence that the conventional interpolation method of determining the background is not always sufficient, particularly for the  situations described above. A more rigorous method is required to deal with such cases of non-structured emission. Motivated by this, and particularly by the approaching solar maximum, we have developed a technique to accurately determine the background.

\section{Fermi-GBM}
\label{sec:fermi}
\begin{figure}
\begin{center}
\begin{tabular}{c}
\includegraphics[scale = 0.5]{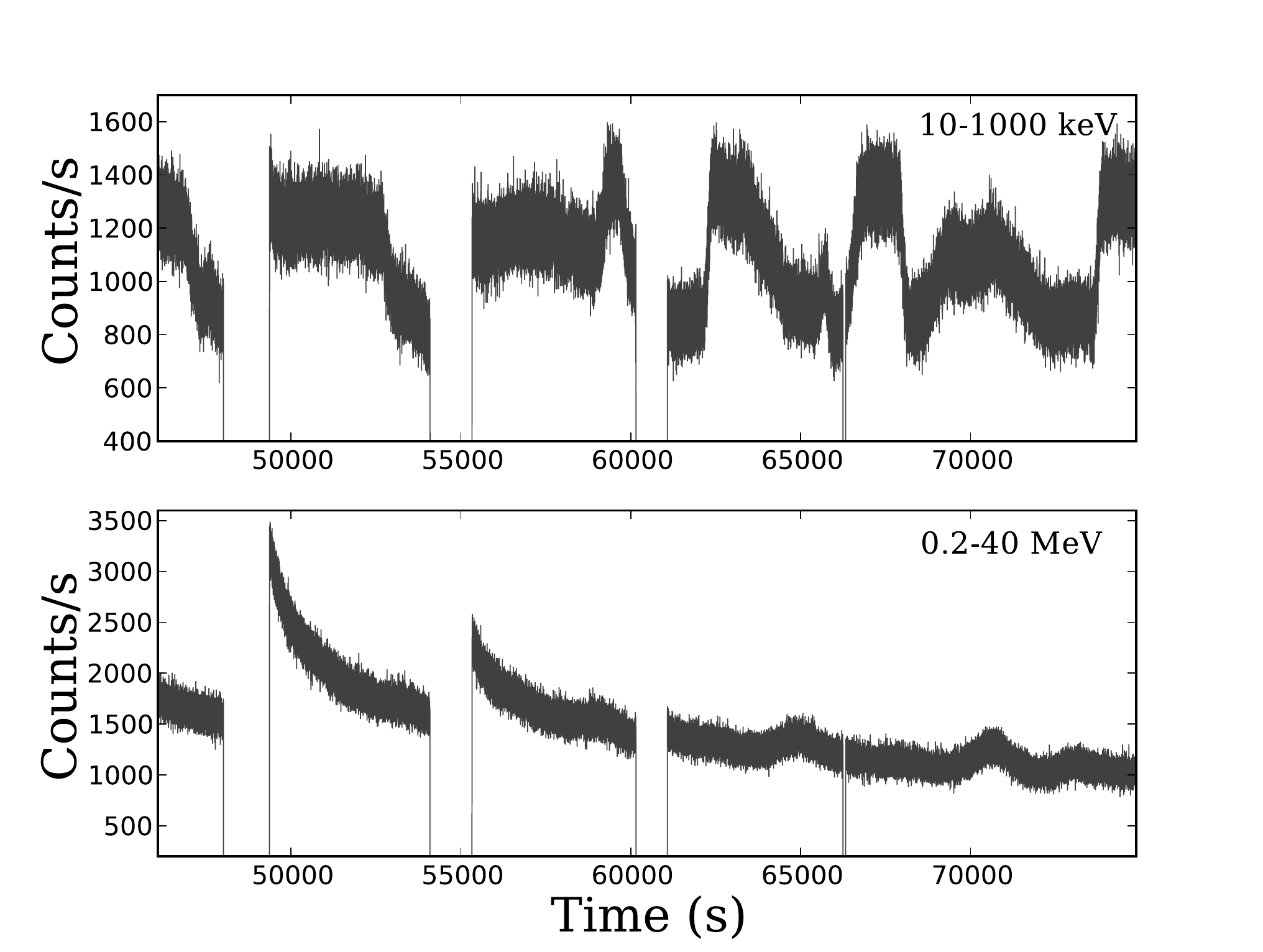}
\end{tabular}
\end{center}
\caption[example] 
{ \label{fig:gbmRates}Background rates in 1.024 s time bins in one NaI detector (top panel) and one BGO detector (bottom panel) over 5 orbits, the time axis is relative to the start of the day. The data gaps correspond to SAA passages. The effect of activation from one such passage can be seen in the increased rates in the BGO directly after exiting the SAA.}
\end{figure}

In orbit since June 2008, \emph{Fermi} has an altitude of $\sim$565 km, inclination of $26^{\circ}$, and period of $\sim$95 minutes.
Both instruments recorded data continuously except when passing through the SAA (see Fig~\ref{fig:orbit}). The primary observation mode of \emph{Fermi} is sky survey mode. This mode optimises the sky coverage of the LAT whilst maintaining near uniform exposure. In this mode the satellite rocks about the zenith such that the entire sky is observed for $\sim$30 minutes every 2 orbits ($\sim$ 3 hours). The rocking causes the satellite pointing to alternate between the northern and southern hemispheres each orbit. The variation in pointing complicates the background in GBM, already highly variable due to the changing geomagnetic conditions. As the satellite slews, bright sources enter/exit the field of view of individual detectors. This manifests in the data as steps, the amplitude of which depends on the strength of the source. The rates in two of the detectors on GBM over 5 orbits ($\sim$25.5 ks) can be seen in Fig.~\ref{fig:gbmRates}.

Of the 14 individual scintillation detectors which make up GBM, 12 are sodium iodide (NaI) which cover the energy range 10 - 1000 keV and 2 are bismuth germanate (BGO) which cover 200 keV - 40 MeV. The 12 NaI detectors are positioned in clusters of three around the spacecraft such that any event which is above the horizon of the Earth's limb will illuminate at least one cluster. 
The two BGO detectors are positioned at opposite sides of the spacecraft, aligned perpendicular to the LAT boresight.

\section{Background estimation using orbital subtraction}
\label{sec:os}
Motivated by Connaughton (2002)~\cite{connaughton_02}, we have implemented a simple method which uses the rates from adjacent days, when the satellite has the same geographical footprint to estimate the background at the time of interest. The orbit of \emph{Fermi} is such that it will be at approximately the same geographical coordinates every 15 orbits ($\sim$24 hrs). However, due to the rocking angle of the spacecraft in sky survey mode, the pointing of the individual GBM detectors is only the same every second orbit. Thus, sources within the field of view (FOV) at $T_0$ may be outside the FOV at $T_0\pm15$ orbits and so this temporal offset cannot be used to approximate the rates at $T_0$. 
This leaves two possibilities; to use the rates from $T_0 \pm 30$ orbits or use the rates from $T_0\pm$14 and $T_0\pm$16 orbits to approximate the rates from $T_0$ $\pm$15 orbits. As the background region is offset in time, the change in the sky background at a given geomagnetic location increases. An initial study showed that the rates from $\pm$30 and $\pm$14,16 orbits can be used interchangeably to estimate the background at the time of interest unless there is a SAA passage exit close to the time of interest, in which case the rates from $\pm$30 should be used\cite{fitzpatrick_11}. 

A major limitation of this technique is that it cannot be employed to investigate times during which the satellite underwent an Autonomous Repoint Request (ARR) or any other deviation from the standard sky survey mode of operation (e.g. ToO). When an ARR is triggered the telescope will slew so that the GBM calculated position is within the LAT FOV. A natural consequence of this is that the periodic pointing is interrupted for the duration of the ARR ($\sim$2.5 hours, formerly 5 hours).   

\section{Verification}
\label{sec:verif}
In order to test the validity of the technique, an initial study was undertaken, the results of which showed good agreement between the estimated and observed rates \cite{fitzpatrick_11}. For four blank fields of duration 2.5 ks, this method generates a background which closely matches the rates in the source region for both NaI and BGO detectors. However, in order to fully validate the method a more rigorous investigation was performed. Specifically, systematic offsets in temporal and spectral properties were investigated and are described below. 
    
\subsection{Blank Sky Tests}\label{sec:blanks}
\begin{figure} \begin{center} \begin{tabular}{c}
\includegraphics[scale = 0.4]{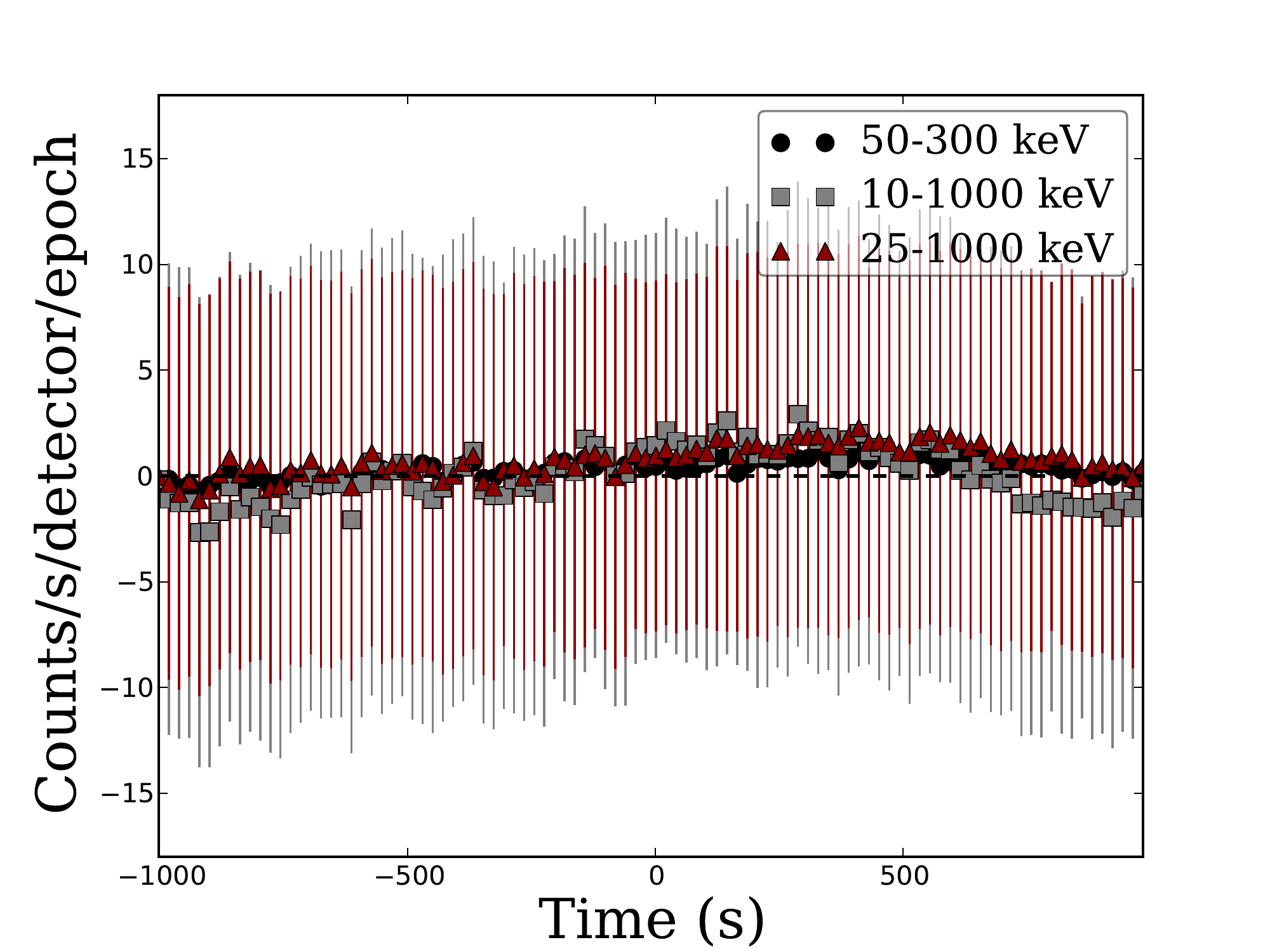} 
\includegraphics[scale = 0.4]{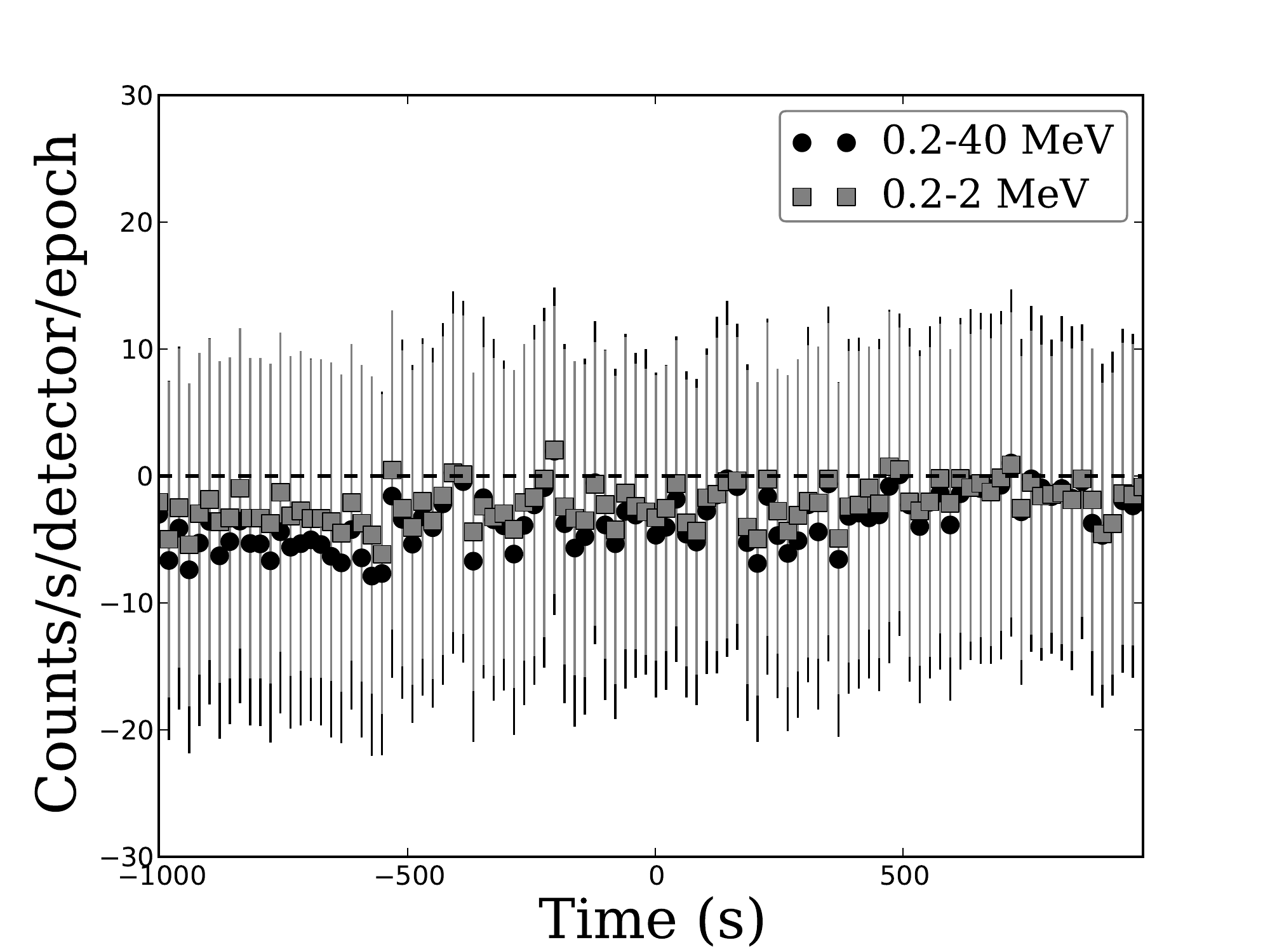}
\end{tabular} \end{center} \caption[example]
{\label{fig:blank_lc}\textbf{Left:} Average lightcurves in each energy band for blank field regions for NaI detectors in $\sim$20 s bins. For each energy band, the data is best described by a $0^{\mbox{th}}$ order polynomial of intercept zero. \textbf{Right:} Average lightcurves in each energy band for blank field regions for BGO detectors in $\sim$20 s bins.  For each energy band, the data exhibits a clear negative bias, however the best fit is a $0^{\mbox{th}}$ order polynomial which is consistent with a straight line of zero intercept.}
\end{figure}

In order to search for systematic offsets, 120 blank fields of duration 2 ks were selected from the first three years of operation (excluding regions of high solar and Soft-Gamma Repeater (SGR) activity). For each region, lightcurves of the \texttt{on} region (defined as the source) and the estimated \texttt{off} region (defined as the background) in three representative energy bands for the NaI and two energy bands for the BGO were generated using the rates from $\pm$30 orbits. Details of these energy bands and the rationale for their selection can be seen in  Table~\ref{tab:eRanges}. The data used were continuous CSPEC, which consist of 128 pseudo-logarithmic spectral bins and 4.096 s timing resolution. Regions where there was clear evidence of an interfering source or a systematic offset between the \texttt{on} and the \texttt{off}  were flagged as bad and discarded. For each energy band, an average residual lightcurve (\texttt{on} $-$ \texttt{off}) was found by averaging all regions and the results are shown in Fig.~\ref{fig:blank_lc}. For each energy band, the average lightcurve is best fit with a $0^{\mbox{th}}$ order polynomial which is consistent with a straight line of intercept zero. For both BGO energy bands, the average lightcurve shows a clear negative bias.

\begin{table}[h] \caption{Energy ranges used in verification process and the rationale for their selection.} 
\label{tab:eRanges} \begin{center}  \begin{tabular}{l l l} 
\hline
\rule[-1ex]{0pt}{3.5ex} Detector Set &  Energy Range (keV) & Rationale \\ \hline
\rule[-1ex]{0pt}{3.5ex} NaI & 10 - 1,000 & Full effective energy range of NaI detectors\\ 
\rule[-1ex]{0pt}{3.5ex} NaI & 25 - 1,000 & Excludes the lowest energy channels of the NaI detectors which\\
& & frequently exhibit  large fluctuations. \\ 
\rule[-1ex]{0pt}{3.5ex} NaI & 50 - 300 & Most sensitive range of the NaI detectors \\ 
\rule[-1ex]{0pt}{3.5ex} BGO & 200 - 40,000 & Full effective energy range of the BGO \\ 
\rule[-1ex]{0pt}{3.5ex} BGO & 200 - 2,000 & Contains the majority of the detected emission in the BGO \\ 
\end{tabular} \end{center} \end{table} 

\begin{figure} \begin{center} \begin{tabular}{c}
\includegraphics[scale = 0.4]{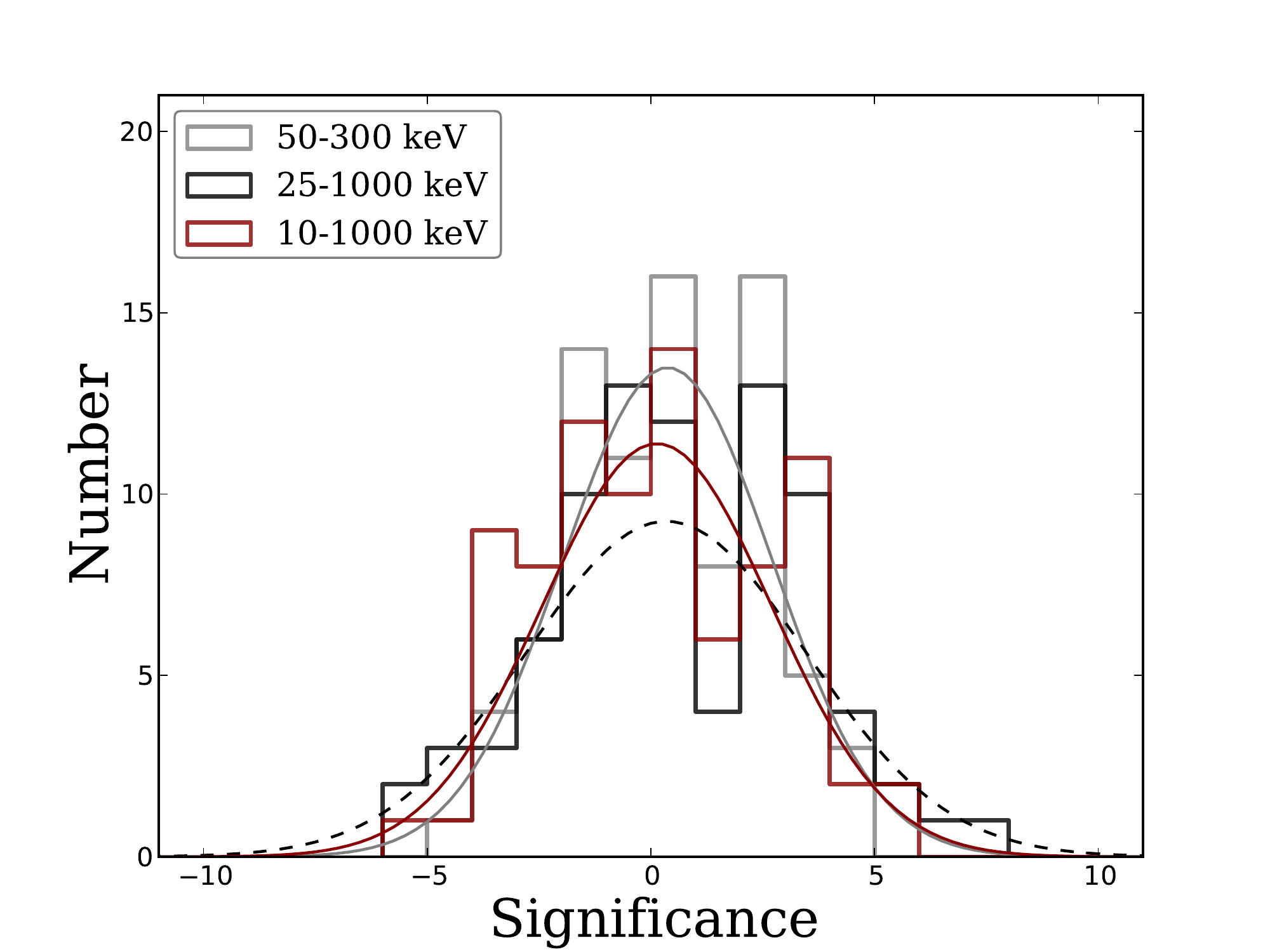}
\includegraphics[scale = 0.4]{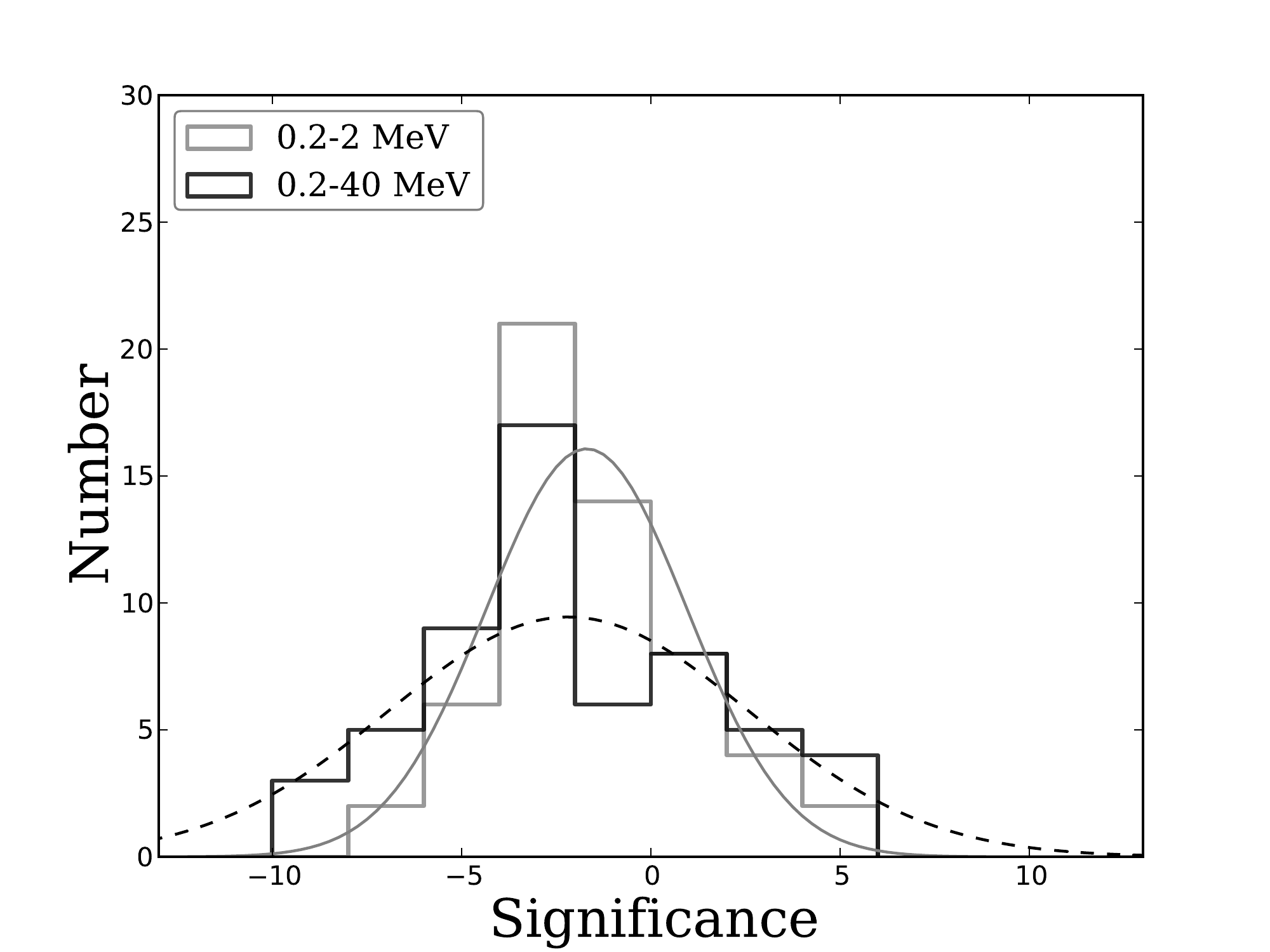}
\end{tabular} \end{center} \caption[example]
{\label{fig:blank_sig}\textbf{Left:} Distribution of significances for blank fields in each NaI energy range. Overplotted are the  results from the modelling of the significances as normal distributions. All three are well described by Gaussian functions: 10-1000 keV and 25-1000 keV have slightly negative and positive means respectively, whereas the mean for 50-300 keV is consistent with zero. \textbf{Right:} Distribution of significances for blank fields in each BGO energy range. The results from the modelling of the significances as normal distributions are overplotted. In both cases, the Gaussian functions which describe the distributions are both wider than the NaI and also both show a clear negative offset, which may be due to the effect of activation following SAA passage.}
\end{figure}

As an additional test, the significance~\cite{li_83} of the \texttt{on} relative to the \texttt{off} was calculated for each region. The distribution of significances for each energy range was then modelled as a normal distribution, the results for the NaI and BGO energy bands can be seen in Fig.~\ref{fig:blank_sig}. For the NaI energy bands, the significances are well-described by Gaussian functions: 10-1000 has a slightly negative mean, 25-1000 has a slightly positive mean and the mean of 50-300 is consistent with zero. This is the expected result, as 50-300 keV is the most sensitive range of the NaI.  For the BGO energy ranges the resulting distributions are a factor of $\sim$2 wider and both exhibit a clear negative offset from zero. This may be due to the effect of activation following SAA passages and warrants further investigation. 

\subsection{Gain Drift}\label{sec:gainDrift}
Both Ackermann et al. (2012) and Briggs et al. (2011)~\cite{2011GeoRL..3802808B, ackermann_12_sfl} report issues with the position of the centroid of the 511 keV line in the spectrum of the BGO, in particular, that spectral fits show that the centroid of the line is in disagreement with the instrumental background line. In light of this, it is important to check whether our method introduces a further uncertainty in the position of spectral features. The current implementation of the orbital subtraction technique aligns the data from the different epochs in channel space and then selects the channel-energy edges from the day of the \texttt{on} region. These edges are also used for the background regions. This assumes that the on-flight calibration is sufficient. 

In orbit, the gain of the detectors can be influenced by a variety of factors, including temperature, ageing of the PMTs, etc. This is corrected on board the spacecraft by the automatic gain control (AGC) which adjusts the high voltage of the detectors to keep a background line in a specific channel~\cite{gbm_paper}. For the NaI detectors the background feature used as reference is the 511 keV annihilation line and in the BGO is the 2.2 MeV neutron capture line. Due to the fact that the high voltage can only be adjusted in discrete steps of 2 V, and to prevent the AGC from chasing statistical fluctuations in the data, the accumulation time used is 90 minutes. The AGC therefore allows small drifts in the gain to proceed uncorrected. In order for the high voltage to be changed, the position of the line must change by 1.4\% in the NaI and 1\% in the BGO. Predictions made pre-launch give a rms gain variation of $<2$\% per detector over one orbit assuming the worst case thermal and magnetic models. 

\begin{figure} \begin{center} \begin{tabular}{c}
\includegraphics[scale = 0.4]{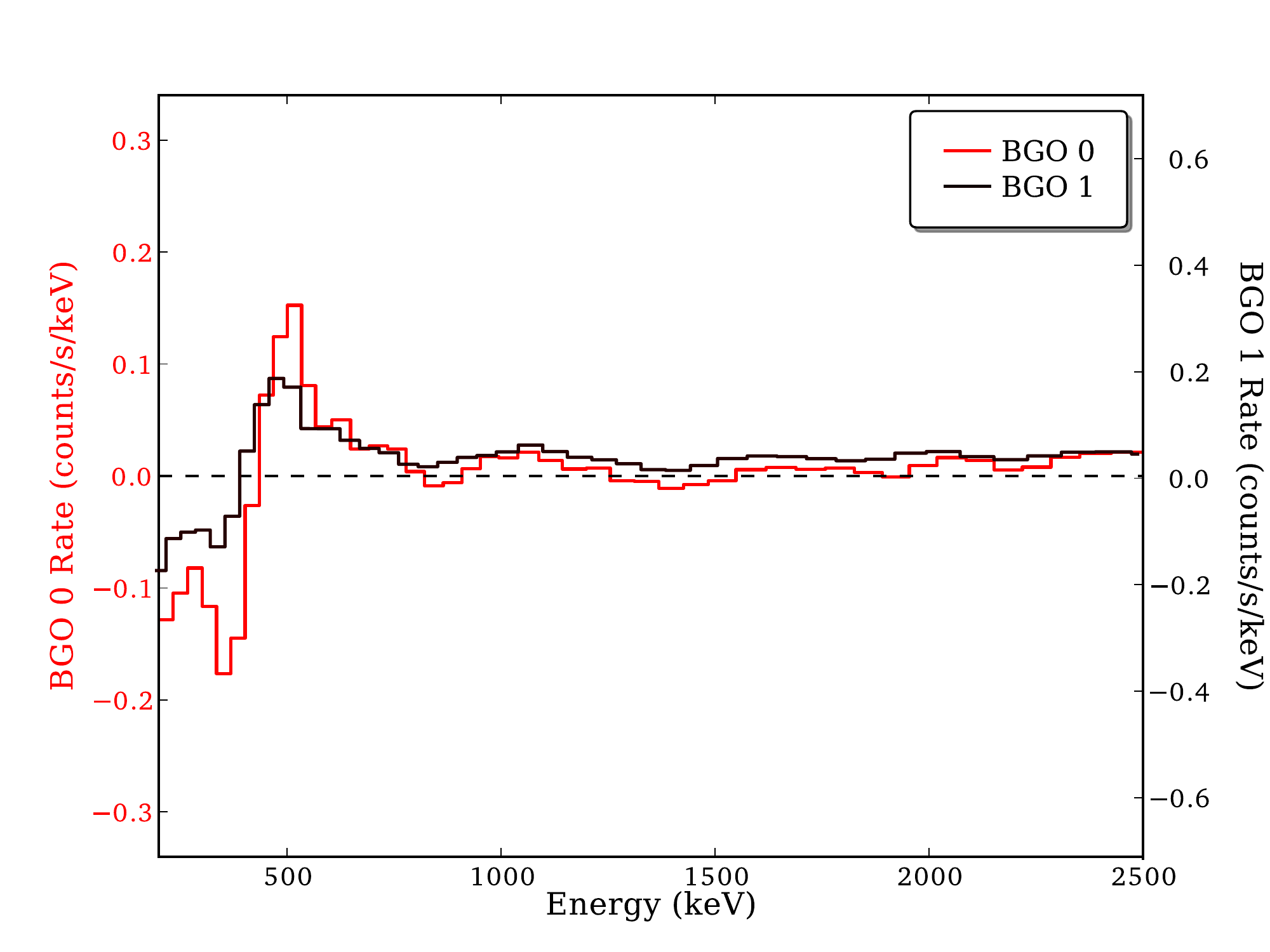}
\includegraphics[scale = 0.4]{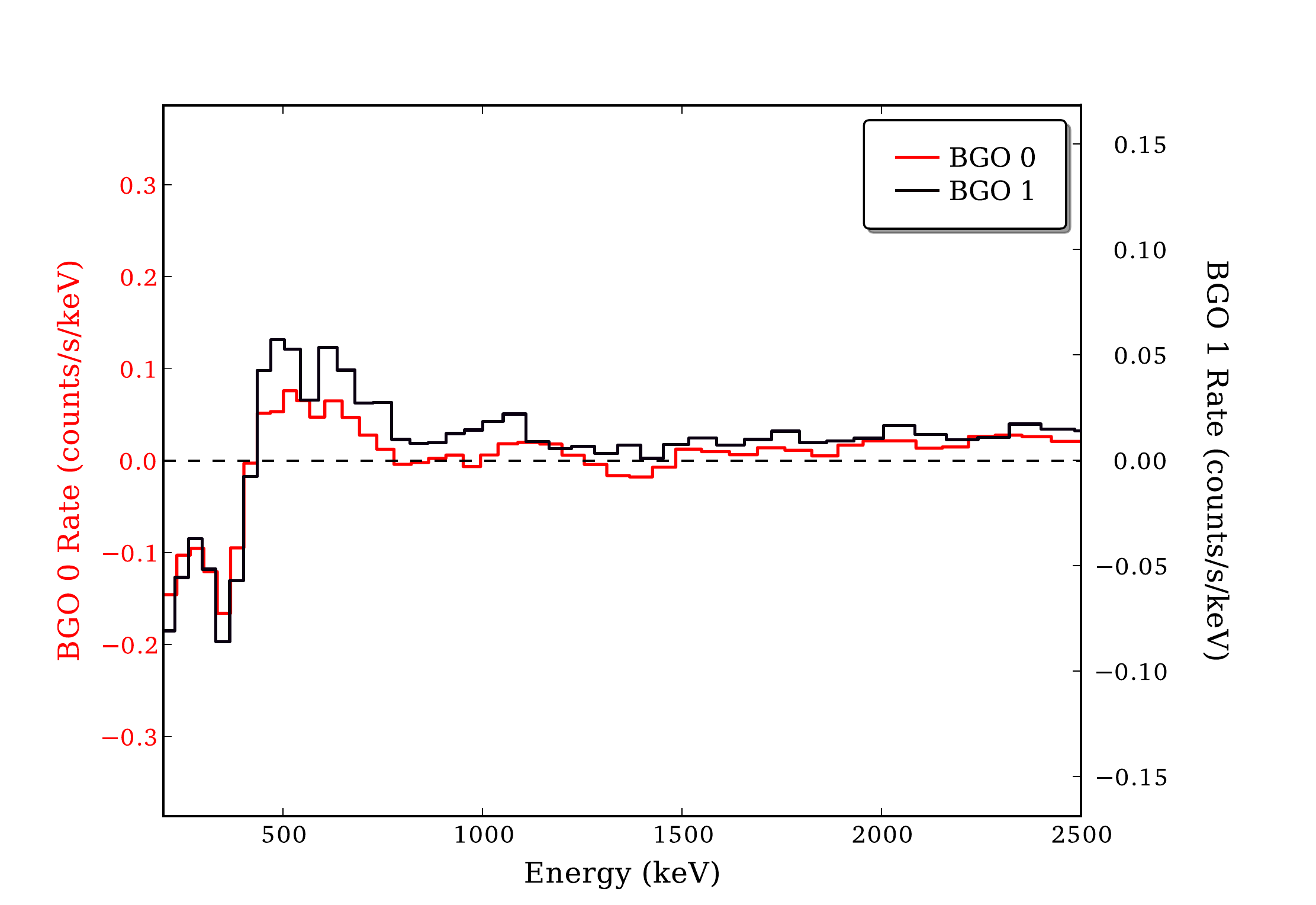}
\end{tabular} \end{center} \caption[example]{ 
\textbf{Left:} Differential count spectra for BGO 0 and 1 for region of low solar activity.
\textbf{Right:} Differential count spectra for BGO 0 and 1 for region of high solar activity. In both cases the residual data around the 511 keV line hints at the possibility of gain drift.
\label{fig:gainDrift}} \end{figure} 

To investigate whether the current implementation is sufficient, two datasets from the BGO detectors of $\sim$3 ks duration separated by 60 orbits ($\sim$324 ks) were selected. The BGO were selected as they have two spectral line features that are visible in the background count spectra (counts/s/keV), the 511 keV and 2.2 MeV lines. The difference between the normalised counts spectra was found and examined for a smearing out of the line features that would be indicative of an uncorrected gain drift. This was performed twice, once for a period of low solar activity in November 2009 and once for high solar activity in March 2012. The results can be seen in Fig.~\ref{fig:gainDrift}. Quantifying the level of gain drift present is complicated by the relative coarseness of the spectral resolution of the detectors on GBM, however examination of the residual plots hints at the possibility of a smearing of the 511 keV line. A lack of statistics limits the amount that can be said for the 2.2 MeV line. In order to draw a more quantitative conclusion on the level of gain drift, a spectral fit to a source with strong line features is required, with a solar flare being the ideal candidate. This was performed for the June 12$^{\mbox{th}}$ 2010 M2 class solar flare~\cite{ackermann_12_sfl}, and is described in the following section (\S~\ref{sec:flare}).

\begin{figure} \begin{center} \begin{tabular}{c}
\includegraphics[scale = .4]{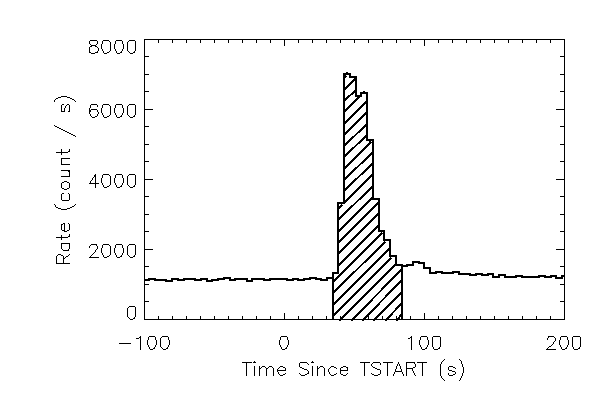}
\includegraphics[scale = .4]{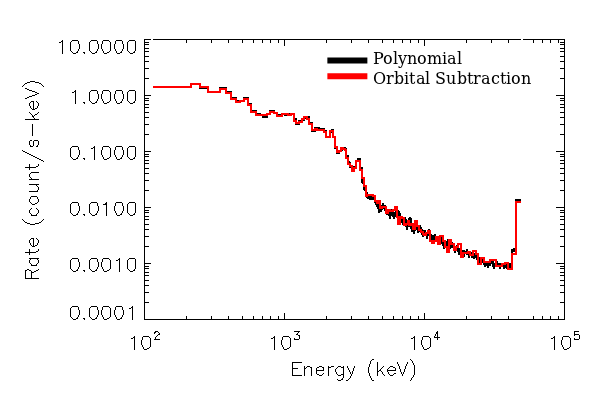}
\end{tabular} \end{center} \caption[example]{ \label{fig:sfl_lc}
\textbf{Left:} BGO 0 lightcurve of June 10$^{\mbox{th}}$ M2 solar flare. The hatched regions indicates the time interval over which the fit was made.
\textbf{Right:} BGO counts spectra showing background from orbital subtraction and background from polynomial fit. The two background estimates agree very well, diverging only slightly at very high energies.}  \end{figure} 

\begin{figure} \begin{center} \begin{tabular}{c}
\includegraphics[scale = 0.4]{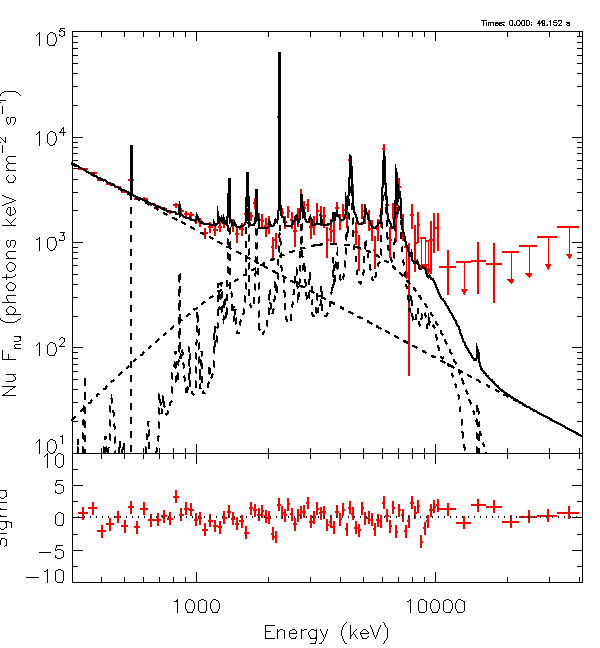}
\includegraphics[scale = 0.4]{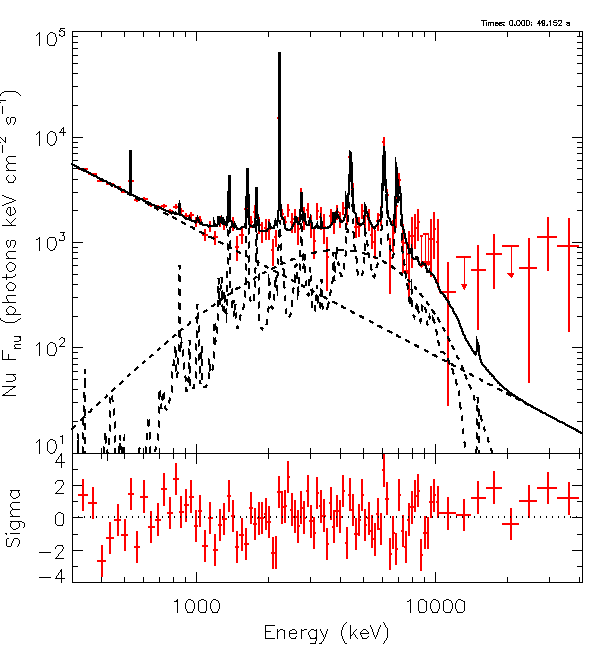}

\end{tabular} \end{center} \caption[example]{ \label{fig:sfl_fit}
\textbf{Left:} $\nu F_\nu$ spectrum for background derived from polynomial interpolation.
\textbf{Right:} $\nu F_\nu$ spectrum for background derived using orbital subtraction technique.
In both cases the spectrum has been fit with a compound model consisting of a power law, power law with exponential cutoff, nuclear de-excitation component and two line profiles at 511 keV and 2.223 MeV.
}  \end{figure}

\subsubsection{Solar Flare June 10th}\label{sec:flare}
On June 12$^{\mbox{th}}$ 2010, both instruments on board \emph{Fermi} detected a GOES M2 class solar flare (SOL2010-06-12T00:57)~\cite{ackermann_12_sfl}. The flare had an impulsive phase of $\sim$50 s and produced gamma-ray emission up to 400 MeV. The short duration of this event implies that the polynomial method should be sufficient and therefore it is an excellent opportunity to compare the results from the two different background methods. The data from the BGO detector on the sun facing side of the satellite were analysed using a background derived with the usual polynomial method and with one derived from the orbital subtraction technique using the rates from $\pm$30 orbits. A lightcurve of the event and a comparison of the two backgrounds can be seen in Fig.~\ref{fig:sfl_lc}. Following Ackermann et al. (2012)~\cite{ackermann_12_sfl}, in each case the spectrum was fit with a compound model consisting of a power law, a power law with exponential cutoff, a nuclear de-excitation component and two Gaussians (511 keV and 2.2 MeV spectral lines). In order to check for a shifting of the spectral line features, the centroid of the 2.2 MeV line was frozen and the 511 keV centroid left as a free parameter. The resulting $\nu F_{\nu}$ spectra for each background model can be seen in Fig.~\ref{fig:sfl_fit}. The fit parameters are comparable for both background models, but the fit statistic increases significantly for the background derived using orbital subtraction ($\Delta$Castor C-Stat\footnote{http://heasarc.gsfc.nasa.gov/docs/xanadu/xspec/XspecManual.pdf}$\sim$60).  For both the polynomial and orbital subtraction methods the centroids of the 511 keV line are consistent ($534 \pm 16$ keV and $530 \pm 17$ keV respectively).

The consistent fit parameters for the different background methods proves that the orbital subtraction method is a viable technique for background estimation. The good agreement of the 511 keV centroid for both interpolated and estimated backgrounds shows that the orbital subtraction method does not introduce an additional uncertainty in the line feature position.

\section{Implementation in a Generalised Background Limited Detector}
The orbital subtraction method described herein can be generalised to any background-limited detector which is in a non-equatorial low earth orbit. The exact implementation of the technique depends on the main observation mode of the satellite. For example, BATSE had a fixed zenith pointing on timescales of 2 weeks, therefore the background could be estimated using data from $\pm$15 orbits, whereas the sky survey mode of \emph{Fermi} means that $\pm$30 orbits must be used. For a generalised instrument, the precise offset depends intrinsically on the orbital parameters of the spacecraft, such as altitude, inclination and orbital period. 

The applicability of the technique depends very much on the energy of the instrument. For hard X-rays (arbitrarily defined as $<25$ keV), the background variation is
composed not only of the local particle flux density, but also includes a contribution from transient X-ray sources. This limits the effectiveness of this technique at these energies, as it cannot account for bright variable sources, which must be manually flagged as bad. In contrast, for gamma-rays, the main variation in the background is due to the local particle flux density (excluding solar contributions). This energy regime is therefore more suited to this technique. An important point is that this method does not account for the contribution of source photons which are scattered by the Earth's upper atmosphere (albedo photons).

\section{Conclusion}
\label{sec:conclusion}
Attempting to study long lived or non-impulsive smooth emission in a background limited instrument like GBM presents many challenges. In order to do so, the orbital background subtraction technique has been developed. This technique has been validated via several tests, which show that it is a viable method of background estimation. 

The best fit to the average residual lightcurve for $\sim$~120 blank fields for 5 energy bands was found to be consistent with zero. The distributions of significances for the 3 NaI energy bands were found to be well described by Gaussians, with means consistent with zero. In contrast, for the two BGO energy bands, the means of the Gaussians are not consistent with zero. This may be due to the effect of activation following SAA passage, and will be the subject of future study. 
 
The June 12$^{\mbox{th}}$ M2 class solar flare was used to compare the results of the conventional polynomial method with that of orbital subtraction. In both cases, the resultant parameters from the spectral analysis are consistent within errors. The 511 keV line centroid and associated error are consistent for both methods. From this, we conclude that the effect of gain drift introduced by the orbital subtraction method is less than the intrinsic gain drift of the instrument and therefore does not contribute significantly to the smearing of spectral features.

The potential of the orbital subtraction technique for studying smooth emission in a background limited instrument has been demonstrated. We are also satisfied that its behaviour is consistent with that of the conventional interpolation method in typical circumstances of impulsive easily-distinguishable emission. 

G.F. acknowledges the support of the Irish Research Council Research for Science, Engineering and Technology. 
\bibliographystyle{spiebib}   
\bibliography{report}   

\begin{thebibliography}{10}

\bibitem{paradijs_97}
{van Paradijs}, J., {Groot}, P.~J., {Galama}, T., {Kouveliotou}, C., {Strom},
  R.~G., {Telting}, J., {Rutten}, R.~G.~M., {Fishman}, G.~J., {Meegan}, C.~A.,
  {Pettini}, M., {Tanvir}, N., {Bloom}, J., {Pedersen}, H.,
  {N{\o}rdgaard-Nielsen}, H.~U., {Linden-V{\o}rnle}, M., {Melnick}, J., {van
  der Steene}, G., {Bremer}, M., {Naber}, R., {Heise}, J., {in't Zand}, J.,
  {Costa}, E., {Feroci}, M., {Piro}, L., {Frontera}, F., {Zavattini}, G.,
  {Nicastro}, L., {Palazzi}, E., {Bennett}, K., {Hanlon}, L., and {Parmar}, A.,
  ``{Transient optical emission from the error box of the {$\gamma$}-ray burst
  of 28 February 1997},'' {\em Nature}~{\bf 386},  686--689 (Apr. 1997).

\bibitem{costa_97}
{Costa}, E., {Frontera}, F., {Heise}, J., {Feroci}, M., {in't Zand}, J.,
  {Fiore}, F., {Cinti}, M.~N., {Dal Fiume}, D., {Nicastro}, L., {Orlandini},
  M., {Palazzi}, E., {Rapisarda\#}, M., {Zavattini}, G., {Jager}, R., {Parmar},
  A., {Owens}, A., {Molendi}, S., {Cusumano}, G., {Maccarone}, M.~C.,
  {Giarrusso}, S., {Coletta}, A., {Antonelli}, L.~A., {Giommi}, P., {Muller},
  J.~M., {Piro}, L., and {Butler}, R.~C., ``{Discovery of an X-ray afterglow
  associated with the {$\gamma$}-ray burst of 28 February 1997},'' {\em
  Nature}~{\bf 387},  783--785 (June 1997).

\bibitem{lat_paper}
{Atwood}, W.~B., {Abdo}, A.~A., {Ackermann}, M., {Althouse}, W., {Anderson},
  B., {Axelsson}, M., {Baldini}, L., {Ballet}, J., {Band}, D.~L.,
  {Barbiellini}, G., and et~al., ``{The Large Area Telescope on the Fermi
  Gamma-Ray Space Telescope Mission},'' {\em Astrophys. J.}~{\bf 697},
  1071--1102 (June 2009).

\bibitem{gbm_paper}
{Meegan}, C., {Lichti}, G., {Bhat}, P.~N., and {et al.}, ``{The Fermi Gamma-ray
  Burst Monitor},'' {\em Astrophys. J.}~{\bf 702},  791--804 (Sept. 2009).

\bibitem{band_93}
{Band}, D., {Matteson}, J., {Ford}, L., {Schaefer}, B., {Palmer}, D.,
  {Teegarden}, B., {Cline}, T., {Briggs}, M., {Paciesas}, W., {Pendleton}, G.,
  {Fishman}, G., {Kouveliotou}, C., {Meegan}, C., {Wilson}, R., and {Lestrade},
  P., ``{BATSE observations of gamma-ray burst spectra. I - Spectral
  diversity},'' {\em Astrophys. J.}~{\bf 413},  281--292 (Aug. 1993).

\bibitem{2009ApJ...706L.138A}
{Abdo}, A.~A., {Ackermann}, M., {Ajello}, M., {Asano}, K., {Atwood}, W.~B.,
  {Axelsson}, M., {Baldini}, L., {Ballet}, J., {Barbiellini}, G., {Baring},
  M.~G., {Bastieri}, D., {Bechtol}, K., {Bellazzini}, R., {Berenji}, B.,
  {Bhat}, P.~N., {Bissaldi}, E., {Blandford}, R.~D., {Bloom}, E.~D.,
  {Bonamente}, E., {Borgland}, A.~W., {Bouvier}, A., {Bregeon}, J., {Brez}, A.,
  {Briggs}, M.~S., {Brigida}, M., {Bruel}, P., {Burgess}, J.~M., {Burrows},
  D.~N., {Buson}, S., {Caliandro}, G.~A., {Cameron}, R.~A., {Caraveo}, P.~A.,
  {Casandjian}, J.~M., {Cecchi}, C., {{\c C}elik}, {\"O}., {Chekhtman}, A.,
  {Cheung}, C.~C., {Chiang}, J., {Ciprini}, S., {Claus}, R., {Cohen-Tanugi},
  J., {Cominsky}, L.~R., {Connaughton}, V., {Conrad}, J., {Cutini}, S.,
  {d'Elia}, V., {Dermer}, C.~D., {de Angelis}, A., {de Palma}, F., {Digel},
  S.~W., {Dingus}, B.~L., {Silva}, E.~d.~C.~e., {Drell}, P.~S., {Dubois}, R.,
  {Dumora}, D., {Farnier}, C., {Favuzzi}, C., {Fegan}, S.~J., {Finke}, J.,
  {Fishman}, G., {Focke}, W.~B., {Fortin}, P., {Frailis}, M., {Fukazawa}, Y.,
  {Funk}, S., {Fusco}, P., {Gargano}, F., {Gehrels}, N., {Germani}, S.,
  {Giavitto}, G., {Giebels}, B., {Giglietto}, N., {Giordano}, F., {Glanzman},
  T., {Godfrey}, G., {Goldstein}, A., {Granot}, J., {Greiner}, J., {Grenier},
  I.~A., {Grove}, J.~E., {Guillemot}, L., {Guiriec}, S., {Hanabata}, Y.,
  {Harding}, A.~K., {Hayashida}, M., {Hays}, E., {Horan}, D., {Hughes}, R.~E.,
  {Jackson}, M.~S., {J{\'o}hannesson}, G., {Johnson}, A.~S., {Johnson}, R.~P.,
  {Johnson}, W.~N., {Kamae}, T., {Katagiri}, H., {Kataoka}, J., {Kawai}, N.,
  {Kerr}, M., {Kippen}, R.~M., {Kn{\"o}dlseder}, J., {Kocevski}, D., {Komin},
  N., {Kouveliotou}, C., {Kuss}, M., {Lande}, J., {Latronico}, L.,
  {Lemoine-Goumard}, M., {Longo}, F., {Loparco}, F., {Lott}, B., {Lovellette},
  M.~N., {Lubrano}, P., {Madejski}, G.~M., {Makeev}, A., {Mazziotta}, M.~N.,
  {McBreen}, S., {McEnery}, J.~E., {McGlynn}, S., {Meegan}, C.,
  {M{\'e}sz{\'a}ros}, P., {Meurer}, C., {Michelson}, P.~F., {Mitthumsiri}, W.,
  {Mizuno}, T., {Moiseev}, A.~A., {Monte}, C., {Monzani}, M.~E., {Moretti}, E.,
  {Morselli}, A., {Moskalenko}, I.~V., {Murgia}, S., {Nakamori}, T., {Nolan},
  P.~L., {Norris}, J.~P., {Nuss}, E., {Ohno}, M., {Ohsugi}, T., {Omodei}, N.,
  {Orlando}, E., {Ormes}, J.~F., {Paciesas}, W.~S., {Paneque}, D., {Panetta},
  J.~H., {Pelassa}, V., {Pepe}, M., {Pesce-Rollins}, M., {Petrosian}, V.,
  {Piron}, F., {Porter}, T.~A., {Preece}, R., {Rain{\`o}}, S., {Rando}, R.,
  {Rau}, A., {Razzano}, M., {Razzaque}, S., {Reimer}, A., {Reimer}, O.,
  {Reposeur}, T., {Ritz}, S., {Rochester}, L.~S., {Rodriguez}, A.~Y., {Roming},
  P.~W.~A., {Roth}, M., {Ryde}, F., {Sadrozinski}, H.~F.-W., {Sanchez}, D.,
  {Sander}, A., {Saz Parkinson}, P.~M., {Scargle}, J.~D., {Schalk}, T.~L.,
  {Sgr{\`o}}, C., {Siskind}, E.~J., {Smith}, P.~D., {Spinelli}, P.,
  {Stamatikos}, M., {Stecker}, F.~W., {Stratta}, G., {Strickman}, M.~S.,
  {Suson}, D.~J., {Swenson}, C.~A., {Tajima}, H., {Takahashi}, H., {Tanaka},
  T., {Thayer}, J.~B., {Thayer}, J.~G., {Thompson}, D.~J., {Tibaldo}, L.,
  {Torres}, D.~F., {Tosti}, G., {Tramacere}, A., {Uchiyama}, Y., {Uehara}, T.,
  {Usher}, T.~L., {van der Horst}, A.~J., {Vasileiou}, V., {Vilchez}, N.,
  {Vitale}, V., {von Kienlin}, A., {Waite}, A.~P., {Wang}, P., {Wilson-Hodge},
  C., {Winer}, B.~L., {Wood}, K.~S., {Yamazaki}, R., {Ylinen}, T., and
  {Ziegler}, M., ``{Fermi Observations of GRB 090902B: A Distinct Spectral
  Component in the Prompt and Delayed Emission},'' {\em Astrophys. J.,
  Lett.}~{\bf 706},  L138--L144 (Nov. 2009).

\bibitem{ackermann_11}
{Ackermann}, M., {Ajello}, M., {Asano}, K., {Axelsson}, M., {Baldini}, L.,
  {Ballet}, J., {Barbiellini}, G., {Baring}, M.~G., {Bastieri}, D., {Bechtol},
  K., {Bellazzini}, R., {Berenji}, B., {Bhat}, P.~N., {Bissaldi}, E.,
  {Blandford}, R.~D., {Bonamente}, E., {Borgland}, A.~W., {Bouvier}, A., and
  et. al, ``{Detection of a Spectral Break in the Extra Hard Component of GRB
  090926A},'' {\em Astrophys. J.}~{\bf 729},  114--+ (Mar. 2011).

\bibitem{1994Natur.372..652H}
{Hurley}, K., {Dingus}, B.~L., {Mukherjee}, R., {Sreekumar}, P., {Kouveliotou},
  C., {Meegan}, C., {Fishman}, G.~J., {Band}, D., {Ford}, L., {Bertsch}, D.,
  {Cline}, T., {Fichtel}, C., {Hartman}, R., {Hunter}, S., {Thompson}, D.~J.,
  {Kanbach}, G., {Mayer-Hasselwander}, H., {von Montigny}, C., {Sommer}, M.,
  {Lin}, Y., {Nolan}, P., {Michelson}, P., {Kniffen}, D., {Mattox}, J.,
  {Schneid}, E., {Boer}, M., and {Niel}, M., ``{Detection of a {$\gamma$}-ray
  burst of very long duration and very high energy},'' {\em Nature}~{\bf 372},
  652--654 (Dec. 1994).

\bibitem{ghisellini_10}
{Ghisellini}, G., {Ghirlanda}, G., {Nava}, L., and {Celotti}, A., ``{GeV
  emission from gamma-ray bursts: a radiative fireball?},'' {\em Mon. Not. R.
  Astron. Soc.}~{\bf 403},  926--937 (Apr. 2010).

\bibitem{kumar_10}
{Kumar}, P. and {Barniol Duran}, R., ``{External forward shock origin of
  high-energy emission for three gamma-ray bursts detected by Fermi},'' {\em
  Mon. Not. R. Astron. Soc.}~{\bf 409},  226--236 (Nov. 2010).

\bibitem{ackermann_12_sfl}
{Ackermann}, M., {Ajello}, M., {Allafort}, A., {Atwood}, W.~B., {Baldini}, L.,
  {Barbiellini}, G., {Bastieri}, D., {Bechtol}, K., {Bellazzini}, R., {Bhat},
  P.~N., {Blandford}, R.~D., {Bonamente}, E., {Borgland}, A.~W., {Bregeon}, J.,
  and et. al, ``{Fermi Detection of {$\gamma$}-Ray Emission from the M2 Soft
  X-Ray Flare on 2010 June 12},'' {\em Astrophys. J.}~{\bf 745},  144 (Feb.
  2012).

\bibitem{connaughton_02}
{Connaughton}, V., ``{BATSE Observations of Gamma-Ray Burst Tails},'' {\em
  Astrophys. J.}~{\bf 567},  1028--1036 (Mar. 2002).

\bibitem{fitzpatrick_11}
{Fitzpatrick}, G., {Connaughton}, V., {McBreen}, S., and {Tierney}, D.,
  ``{Uncovering low-level Fermi/GBM emission using orbital background
  subtraction},'' {\em ArXiv e-prints}  (Nov. 2011).

\bibitem{li_83}
{Li}, T.-P. and {Ma}, Y.-Q., ``{Analysis methods for results in gamma-ray
  astronomy},'' {\em Astrophys. J.}~{\bf 272},  317--324 (Sept. 1983).

\bibitem{2011GeoRL..3802808B}
{Briggs}, M.~S., {Connaughton}, V., {Wilson-Hodge}, C., {Preece}, R.~D.,
  {Fishman}, G.~J., {Kippen}, R.~M., {Bhat}, P.~N., {Paciesas}, W.~S.,
  {Chaplin}, V.~L., {Meegan}, C.~A., {von Kienlin}, A., {Greiner}, J., {Dwyer},
  J.~R., and {Smith}, D.~M., ``{Electron-positron beams from terrestrial
  lightning observed with Fermi GBM},'' {\em Geophys. Res. Lett.}~{\bf 38},
  2808 (Jan. 2011).

\end{thebibliography}

\end{document}